\documentclass[graybox]{svmult}
\usepackage{wrapfig}
\usepackage{hyperref}
\usepackage{mathptmx}       
\usepackage{helvet}         
\usepackage{courier}        
\usepackage{type1cm}        
\usepackage{makeidx}         
\usepackage{graphicx}        
\usepackage{multicol}        
\usepackage[bottom]{footmisc}

\usepackage{float}
\usepackage{amsmath}
\usepackage{amssymb}
\usepackage{algorithm}
\usepackage{algorithmic}
\usepackage{graphics}
\usepackage{graphicx}
\usepackage{epsfig}
\usepackage{url}
\usepackage{multirow}
\usepackage{xspace}
\usepackage{threeparttable}
\usepackage{comment}
\usepackage{multicol}
\usepackage{physics}
\usepackage{esvect}
\usepackage{framed}
\usepackage{tikz}
\usetikzlibrary{shapes.geometric, arrows, positioning}
\usepackage{colortbl}
\usepackage{multirow}

\tikzstyle{box} = [rectangle, minimum width=3cm, minimum height=1.5cm, text centered, draw=black, fill=blue!20]
\tikzstyle{channel} = [thick,->,>=stealth]
\tikzstyle{classical} = [thick,<->,>=stealth]
\tikzstyle{encryptor} = [rectangle, minimum width=2cm, minimum height=1cm, text centered, draw=black, fill=gray!20]
\tikzstyle{kms} = [rectangle, minimum width=2cm, minimum height=1cm, text centered, draw=black, fill=red!20]
\tikzstyle{message} = [rectangle, minimum width=1cm, minimum height=1cm, text centered, draw=black, fill=yellow!20]
\tikzstyle{qkd_system} = [ellipse, draw=black, dashed, fill=orange!20, minimum width=10cm, minimum height=5cm]

\begin{document}

\title{Quantum Key Distribution}
\author{Sebastian Kish, Josef Pieprzyk \& Seyit Camtepe}

\institute{{Dr. Seyit Camtepe} \at{CSIRO, Australia}, \href{mailto:Seyit.Camtepe@data61.csiro.au}{Seyit.Camtepe@data61.csiro.au},
\\
{Dr. Sebastian Kish} \at{CSIRO, Australia}, \href{mailto:Sebastian.Kish@data61.csiro.au}{Sebastian.Kish@data61.csiro.au}
\\
{Prof. Josef Pieprzyk} \at{Institute of Computer Science, Polish Academy of Sciences, Poland and CSIRO, Australia},
\href{mailto: Josef.Pieprzyk@data61.csiro,au}{Josef.Pieprzyk@data61.csiro,au}
}

\maketitle
\label{label}


Quantum Key Distribution (QKD) is a technology that ensures secure communication by leveraging the principles of quantum mechanics, such as the no-cloning theorem and quantum uncertainty. This chapter provides an overview of this quantum technology's maturity and trends. It highlights significant advancements in single-photon sources and detection technologies that have brought QKD closer to widespread adoption, including real-world deployments by industry leaders. While addressing challenges such as cost, integration, standardization, and the need for quantum repeaters, the chapter emphasizes the growing importance of QKD in securing mission-critical communications against future quantum threats. Through its unique ability to achieve information-theoretic security, QKD is poised to play a vital role in quantum-safe cryptographic algorithms and protocols.

\section{Introduction}
Recent advancements have led to the development of a new class of computers known as quantum computers, which possess the potential to break many of the encryption algorithms currently in use. To counteract this threat, Quantum Key Distribution (QKD) has emerged as a pivotal quantum technology, offering a fundamentally secure method of communication that does not rely on assumptions about an adversary's computational power. Globally, efforts are underway to connect individual point-to-point QKD links into expansive testbed networks, bringing us closer to commercially viable solutions. While countries like the USA, China, and Japan are leading in advancing QKD technology through large-scale research and development initiatives, Switzerland has distinguished itself in the commercialization of QKD, leveraging its strong foundation in both academic research and industrial expertise. By addressing the remaining challenges, there's an opportunity for international industry and government to play a prominent role in the secure communication landscape of the future.

\subsection{Challenges and Opportunities}

Quantum computers will soon threaten secure data traffic, necessitating new cryptographic methods. Current cryptography relies on symmetric encryption (e.g., AES) and public-key encryption, with the latter often used to distribute symmetric keys. While symmetric encryption is less vulnerable, Grover's algorithm weakens AES-256 to a 128-bit security level. However, public-key methods like RSA are significantly more at risk due to Shor's algorithm, which provides an exponential speedup for breaking these systems. This highlights the need for quantum-safe solutions, such as Quantum Key Distribution (QKD), which leverages quantum mechanics to securely exchange symmetric keys and detect eavesdropping, making it immune to quantum and classical computational advances. Unlike post-quantum cryptography (PQC), which relies on unproven assumptions about mathematical problem hardness, QKD provides a future-proof solution, mitigating the risk of “store now, decrypt later” attacks and ensuring long-term data confidentiality without dependency on computational assumptions.
\begin{figure}[th!]
\caption{Quantum Key Distribution (QKD) system}
\begin{tikzpicture}[node distance=3cm]

\node (qkd_system) [qkd_system, minimum width=10.5cm, minimum height=3cm] {};

\node (alice) [box, left=1.2cm of qkd_system.center] {Alice};
\node (bob) [box, right=1.2cm of qkd_system.center] {Bob};

\draw [channel, red, thick] (alice.east) -- node[midway, below] {Quantum Channel} (bob.west);
\draw [classical, black,thick] (alice.north east) -- ++(1.75cm,0) -- ++(-1.0cm,0) node[midway, below] {Classical Channel} -- (bob.north west);

\node[right=-3cm,below=-.5cm of alice] {$K_A$};
\node[right=3cm,below=-.5cm of bob] {$K_B$};
\node[below=0.9cm] {Shared Key $K_A=K_B=K$};
\node (alice_enc) [encryptor, below=3.5cm of alice] {Encryptor};
\node (bob_enc) [encryptor, below=3.5cm of bob] {Encryptor};

\node (alice_kms) [kms, below=1.5cm of alice] {KMS};
\node (bob_kms) [kms, below=1.5cm of bob] {KMS};
\draw[thick,<->,>=stealth] (alice_kms.south)--(alice_enc.north);
\draw[thick,<->,>=stealth] (bob_kms.south)--(bob_enc.north);
\draw[thick,-,>=stealth] (alice_kms.east)--(bob_kms.west);
\draw [channel] (alice.south) -- (alice_kms.north);
\draw [channel] (bob.south) -- (bob_kms.north);

\node[below=-1.5cm] {QKD};
\node[below=2.2cm] {KMS Link};
\node[below=4.2cm] {Encrypted Message};
\draw [channel] (alice_enc.east) -- (bob_enc.west);
\draw [channel] (bob_enc.west) -- (alice_enc.east);

\node (alice_message) [message, left=0.5cm of alice_enc] {Message};
\node (bob_message) [message, right=0.5cm of bob_enc] {Message};

\draw [channel] (alice_message.east) -- (alice_enc.west);
\draw [channel] (alice_enc.west)--(alice_message.east);

\draw [channel] (bob_message.west) -- (bob_enc.east);
\draw [channel] (bob_enc.east)--(bob_message.west);
\end{tikzpicture}
\label{qkd}
\end{figure}
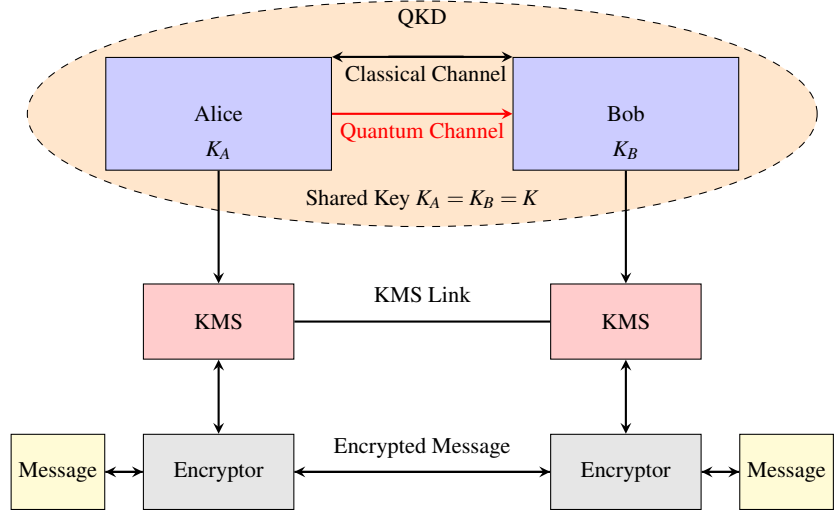
\subsection{Definition of QKD} 
All QKD protocols are executed by two parties Alice and Bob as depicted in Fig. \ref{qkd}.
Their goal is to establish a common and secret key $K$.
An adversary, Eve, is assumed to have access to communication channels used by them.
Alice and Bob are connected by both unidirectional quantum and bidirectional classical channels.
The {\it quantum channel} can be optical fiber or alternatively free space, which is able to transmit photons.
The {\it classical channel} is assumed to be authenticated, i.e. a receiver is able to verify if a message comes
from an alleged sender. It can be implemented by appending a digital signature or message authentication code (MAC)
 to the message. A key management system (KMS) is used to manage keys that can also function as a standards based key scheduler for shared randomness. QKD's security is rooted in two fundamental principles of quantum mechanics: the \emph{no-cloning} theorem and the quantum \emph{uncertainty} principle.

The no-cloning theorem states that it is impossible to create an exact copy of an arbitrary unknown quantum state. This principle underpins the security of QKD because it prevents an eavesdropper like Eve from intercepting photons sent by Alice to Bob and duplicating them to avoid detection. Any attempt by Eve to measure or clone the quantum states will inevitably disturb them, introducing detectable anomalies.

Additionally, the quantum uncertainty principle ensures that certain pairs of properties (e.g., position and momentum, or orthogonal polarization states) cannot be simultaneously measured with perfect accuracy. In the context of QKD, if Eve attempts to intercept and measure the quantum states sent by Alice, her actions will disturb the states in a way that introduces errors into the key generation process. Bob can detect these disturbances when comparing a subset of their measurement results with Alice through an authenticated public channel. If the error rate exceeds a predefined threshold, Alice and Bob know the communication has been compromised and can discard the affected key.

Through this process, Alice and Bob can ensure that their key is secure, even in the presence of a potential eavesdropper, provided they have an authenticated communication channel for exchanging classical information.

\section{Maturity of QKD Technology}

\subsection{QKD Systems}
The development of QKD systems has reached a level of technical maturity, with multiple vendors producing commercially available products tailored for various applications. Companies like ID Quantique, Toshiba, QuintessenceLabs, and LuxQuanta are leading efforts to commercialize QKD, offering solutions that integrate seamlessly into existing communication infrastructures. These vendors provide systems based on diverse protocols, such as decoy-state BB84, Gaussian-modulated CV-QKD, and Coherent One-Way QKD, each optimized for different use cases, as summarized in Table \ref{tab:qkd_security_compact_updated}.

\begin{table}[ht]
\centering
\caption{QKD Protocol Security, Implementation Maturity, and Vendors}
\label{tab:qkd_security_compact_updated}
\resizebox{\textwidth}{!}{\begin{tabular}{|p{3cm}|p{2.3cm}|p{1.8cm}|p{1.8cm}|p{3.3cm}|}
\hline
\textbf{Protocol}                & \textbf{Aspect}          & \textbf{Current}       & \textbf{Future Outlook} & \textbf{Vendors}               \\ \hline
\multirow{2}{=}{Decoy-State (includes BB84)} 
                                 & Protocol Security        & \cellcolor{green!20}Proven               & \cellcolor{green!20}Stable     & \multirow{2}{=}{ID Quantique, Toshiba, ThinkQuantum} \\ \cline{2-4}
                                 & Implementation Maturity  & \cellcolor{green!20}Promising             & \cellcolor{green!20}Mature  & \\ \hline
\multirow{2}{=}{Gaussian-Modulated CV-QKD} 
                                 & Protocol Security        & \cellcolor{green!20}Proven          & \cellcolor{green!20}Stable  & \multirow{2}{=}{QuintessenceLabs, LuxQuanta} \\ \cline{2-4}
                                 & Implementation Maturity  & \cellcolor{green!20}Improving             & \cellcolor{green!20}Mature  & \\ \hline
\multirow{2}{=}{Discrete-Modulated CV-QKD (e.g., QPSK)} 
                                 & Protocol Security        & \cellcolor{yellow!20}Developing               & \cellcolor{green!20}Promising & \multirow{2}{=}{Huawei, AIT} \\ \cline{2-4}
                                 & Implementation Maturity  & \cellcolor{yellow!20}Moderate    & \cellcolor{green!20}Improving & \\ \hline
\multirow{2}{=}{Coherent One-Way} 
                                 & Protocol Security        & \cellcolor{yellow!20}Developing          & \cellcolor{green!20}Promising  & \multirow{2}{=}{ID Quantique, QNu Labs} \\ \cline{2-4}
                                 & Implementation Maturity  & \cellcolor{yellow!20}Moderate             & \cellcolor{green!20}Improving  & \\ \hline
\multirow{2}{=}{Entanglement-Based Protocols (e.g., E91)} 
                                 & Protocol Security        & \cellcolor{green!20}Proven               & \cellcolor{green!20}Stable     & \multirow{2}{=}{S-Fifteen, Toshiba, ID Quantique} \\ \cline{2-4}
                                 & Implementation Maturity  & \cellcolor{red!20}Challenging            & \cellcolor{yellow!20}Developing & \\ \hline
\multirow{2}{=}{Twin-Field QKD} 
                                 & Protocol Security        & \cellcolor{green!20}Promising         & \cellcolor{green!20}Advancing  & \multirow{2}{=}{Toshiba demo (not yet available)} \\ \cline{2-4}
                                 & Implementation Maturity  & \cellcolor{red!20}Challenging            & \cellcolor{yellow!20}Developing & \\ \hline
\end{tabular}}
\end{table}

Over the past two decades, advancements in single-photon sources and detection technologies have significantly reduced costs, making QKD more accessible. In particular, the development and widespread adoption of avalanche photodiodes (APDs) for the decoy-state QKD protocol have eliminated the reliance on costly superconducting nanowire single-photon detectors (SNSPDs), which require cryogenic cooling. For example, Toshiba's proprietary T12 protocol leverages APDs and other cost-effective single-photon technologies to achieve key distribution over distances of up to 150 km \cite{Lucamarini:13}. These innovations are crucial in reducing the cost barriers associated with QKD systems, enabling their deployment in more affordable and scalable configurations, as reflected in the advancements noted in Table \ref{tab:qkd_security_compact_updated}.

Other approaches to reduce costs and enhance compatibility with existing optical communication systems include Continuous-Variable QKD (CV-QKD). QuintessenceLabs Inc., an Australian company, has released a product based on the GG02 protocol and heterodyne detection. These protocols, while less expensive compared to discrete-variable QKD systems, are limited in range due to phase-locking noise. Similarly, LuxQuanta has introduced a CV-QKD system available through the AWS Marketplace, demonstrating growing commercial interest in this cost-effective approach to quantum-secure communication.

To further reduce production costs, ID Quantique has developed a product based on the Coherent One-Way QKD protocol. Although this protocol currently lacks a fully proven information-theoretic security proof, it leverages off-the-shelf components to provide a more practical and scalable solution. Such advancements make quantum communication systems increasingly accessible to a broader range of users, particularly for enterprise applications.

These QKD protocol developments, as summarized in Table \ref{tab:qkd_security_compact_updated}, illustrate the ongoing progress in making QKD systems more affordable, scalable, and adaptable to existing communication infrastructure, driving broader adoption across industries.
\begin{table}[th]
\centering
\resizebox{\textwidth}{!}{\begin{tabular}{|p{2cm}|p{1.5cm}|p{1.2cm}|p{5cm}|p{1.2cm}|p{1cm}|p{1cm}|p{1.5cm}|}
\hline
\textbf{QKD Network} & \textbf{Region} & \textbf{Year Commissioned} & \textbf{Key Features} & \textbf{Maturity} & \textbf{Number of Nodes} & \textbf{Covered Distance} & \textbf{Use} \\
\hline
\multirow{2}{*}{MadQCI} & \multirow{2}{*}{Spain} & \multirow{2}{*}{2021} & Integrated with commercial telecom networks; compatible with IPsec encryption devices; utilizes ID Quantique, Toshiba, AIT \& Huawei QKD systems \cite{madqci} & \multirow{2}{*}{High} & 10 & 200 km & Commercial \\
 & & & SDN architecture implemented & & & & \\
\hline
\multirow{2}{*}{SK Telecom} & \multirow{2}{*}{South Korea} & \multirow{2}{*}{2019} & Nationwide deployment for government organizations; subscription-based QKD service for enterprises; employs ID Quantique's QKD systems \cite{sktelecom} & \multirow{2}{*}{High} & 15 & 150 km & Commercial \\
 & & & SDN-based control of heterogeneous QKD networks & & & & \\
\hline
\multirow{2}{*}{Singapore QKD} & \multirow{2}{*}{Singapore} & \multirow{2}{*}{2020} & Integrated into national infrastructure for secure communication; positioned as a regional hub for quantum security; collaborates with ID Quantique & \multirow{2}{*}{High} & 8 & 100 km & Commercial \\
 & & & & & & & \\
\hline
\multirow{2}{*}{EuroQCI} & \multirow{2}{*}{EU} & \multirow{2}{*}{2023} & Developing a quantum network across 27 member states; focus on security for critical infrastructures; involves multiple vendors including Toshiba and ID Quantique, cross-border space links, intra-city and inter-city fibre links& \multirow{2}{*}{High} & $2-10$ & 10-1000 km& Research \\
 & & & & & & & \\
\hline
PSNC QKD Link& \multirow{2}{*}{Poland} & \multirow{2}{*}{2022} & 380 km intercity QKD link within PIONIER network; includes international QKD links with the Czech Republic; utilizes ID Quantique's systems & \multirow{2}{*}{High} & 5 & 380 km & Research \\
 & & & & & & & \\
\hline
\multirow{2}{*}{Cambridge QKD} & \multirow{2}{*}{UK} & \multirow{2}{*}{2023} & Operates on dense wavelength division multiplexing (DWDM) networks; demonstrates high-bandwidth quantum communication; vendor information not specified & \multirow{2}{*}{High} & 3 & 25 km & Research \\
 & & & & & & & \\
\hline
DARPA Quantum Network & \multirow{2}{*}{USA} & \multirow{2}{*}{2004} & Integrated with Internet technologies; used QKD-derived keys for IPsec; one of the first QKD networks deployed; utilized proprietary QKD systems & \multirow{2}{*}{High} & 3 & 50 km & Research \\
 & & & & & & & \\
\hline
Bristol Quantum Network& \multirow{2}{*}{UK} & \multirow{2}{*}{2020} & QKD provided over 5GUK test network using specially developed Open Source software, also trusted-node free quantum network; University of Bristol & \multirow{2}{*}{High} & 4-8 & 13 km & Research \\
 & & & & & & & \\
 \hline
Tokyo QKD Network& \multirow{2}{*}{Japan} & \multirow{2}{*}{2010} & Multi-node testbed on NICT’s JGN-X open fiber network; collaborative research platform for universities and industry; involved multiple vendors & \multirow{2}{*}{Medium} & 7 & 300 km & Research \\
\hline
\multirow{2}{*}{CSIRO Testbed} & \multirow{2}{*}{Australia} & \multirow{2}{*}{2024} & Laboratory-based QKD research environment; focuses on experimental validation and development; employs QuintessenceLabs' QKD system & \multirow{2}{*}{Low} & 2 & 20 km & Research\\
 & & & & & & & \\
\hline
\end{tabular}}
\caption{Maturity Levels and Vendors of QKD Activities and Testbeds Worldwide}
\label{table:qkd_maturity_vendors}
\end{table}
\subsection{QKD Activities and Testbeds}
QKD activities have advanced significantly, transitioning from purely experimental setups to more sophisticated testbeds and early-stage deployments. Some of the most notable QKD initiatives demonstrating significant progress are shown in Table \ref{table:qkd_maturity_vendors}. A notable example is the SwissQuantum testbed in Geneva, launched in 2008. Spanning approximately 20 kilometers, it connected multiple nodes, including corporate offices and data centers, serving as a robust platform for evaluating QKD technology \cite{stucki}. Such projects highlight the potential for integrating QKD into modern communication systems and pave the way for broader adoption.

The Madrid Quantum Communication Infrastructure (MadQCI) demonstrates significant progress in QKD by integrating quantum communication channels with classical channels for data transmission and network control, managed dynamically through Software-Defined Networking (SDN) \cite{madqci}. Its architecture includes a Local Key Management System (LKMS) that collects, stores, and manages keys from QKD modules, enabling real-time network monitoring and dynamic reconfiguration. By addressing challenges in hybrid network management, MadQCI highlights the feasibility of scalable QKD systems for real-world applications.

In South Korea, SK Telecom, in partnership with ID Quantique, has developed one of the most advanced QKD testbeds globally, deploying QKD systems over the past five years to connect 48 government organizations \cite{sktelecom}. This testbed secures critical communications for government, financial institutions, and enterprises, showcasing the scalability of quantum-safe solutions. Additionally, QKD services have been successfully deployed at Equinix’s SL1 data center, offering enterprise clients a subscription-based model that reduces upfront costs, demonstrating the practicality of large-scale QKD implementations.

Singapore has also made significant strides in quantum communication by building a comprehensive QKD testbed in collaboration with ID Quantique. As part of its nationwide quantum security initiative, Singapore has deployed QKD technology to secure its sensitive government and enterprise communications, positioning itself as a leader in quantum-safe communication in Asia. This effort integrates QKD into the broader national infrastructure, demonstrating its commitment to securing critical communications against future quantum threats. With these developments, Singapore is poised to be a hub for quantum innovation in the region.

The European Union’s EuroQCI initiative is building a secure quantum communication infrastructure across all 27 EU Member States, including Poland, to enhance security for critical infrastructures and government institutions. As part of this effort, Poland has been advancing its QKD activities through the Poznań Supercomputing and Networking Center (PSNC) in collaboration with ID Quantique, establishing a 380 km intercity QKD link between Poznań and Warsaw within the PIONIER network, and creating the first international QKD link with Czech institutions between Cieszyn and Ostrava. These efforts position Poland as a key contributor to EuroQCI, integrating quantum technologies into secure communication testbeds.

These efforts in South Korea and Singapore, alongside initiatives in Europe with EuroQCI and MadQCI, underscore the global momentum toward quantum-safe communication. They highlight the potential of QKD to transition from isolated demonstrations to integral components of national and enterprise-level cybersecurity strategies.



\section{Trends and Innovations of QKD}
QKD is rapidly advancing through theoretical and practical innovations, offering information-theoretic security based on quantum mechanics. While foundational protocols like BB84 and decoy-state QKD have established security proofs, newer protocols often lack complete analyses, particularly under real-world conditions with finite datasets. Research is focused on addressing these gaps to ensure practical security.

In parallel, challenges such as implementation vulnerabilities, cost, scalability, standardization, and integration with existing systems continue to hinder large-scale deployment. Despite these hurdles, significant progress is being made, as outlined below:

\begin{itemize}

\item[\textbf{I.}] \textbf{Implementation Security}

QKD’s theoretical promise of ``unconditional security" can be compromised in real-world implementations due to hardware imperfections. These vulnerabilities have been exploited in various \textit{side-channel attacks} and/or quantum hacking, such as photon-number-splitting (PNS) attacks, detector blinding, and time-shift attacks \cite{makarov}.

\emph{Severity}: Medium. While vulnerabilities exist, countermeasures like measurement-device-independent QKD (MDI-QKD) and (semi-) device-independent QKD are advancing rapidly and already offer solutions for mitigating these risks.

\emph{Timeline for Resolution}: Short to medium term (3–7 years). Many countermeasures are being standardized and are expected to integrate seamlessly into commercial systems soon.

\item[\textbf{II.}] \textbf{Limited Role as a Cryptographic Solution}

QKD is often criticized for being a partial solution, as it generates keying material but does not inherently provide source authentication. The authentication of the QKD transmission source typically relies on pre-placed symmetric keys or asymmetric cryptography \cite{nsa}, which limits its standalone utility.

Emerging quantum technologies, such as Quantum Digital Signatures (QDS) and Quantum-Secure Identifiers (QSIs), leverage quantum principles to address this limitation. Hybrid solutions combining QKD with post-quantum cryptography (PQC), such as lattice-based cryptographic algorithms, are also gaining traction as a practical approach for robust security.

\emph{Severity}: Medium. Current cryptographic tools and emerging technologies provide adequate solutions, making this a manageable challenge.

\emph{Timeline for Resolution}: Short term (1–3 years) for hybrid solutions; longer term (5–10 years) for full reliance on quantum-based authentication technologies like QDS and QSIs.

\item[\textbf{III.}] \hspace{0.05cm} \textbf{Key Extraction Efficiency}

Efficiently extracting secret keys from raw quantum measurement data is critical for real-time operation. The bottleneck in error reconciliation, especially under noisy conditions or high-loss scenarios, has been a challenge. However, modern low-leakage error correction codes and advanced reconciliation techniques already perform well, with minimal delays in key generation.

\emph{Severity}: Low. While not ideal in all scenarios, backlogged keys can be stored and processed without compromising security.

\emph{Timeline for Resolution}: Very short term (1–2 years). Existing solutions are already effective and are continuously improving with incremental advancements in algorithms and hardware acceleration.

\item[\textbf{IV.}] \hspace{0.05cm}\textbf{Cost and Scalability}

The cost of deploying QKD infrastructure, particularly for discrete-variable (DV-QKD) systems, remains a barrier due to the specialized hardware required. Continuous-variable (CV-QKD) systems, which are more cost-effective and compatible with standard telecom components, face limitations in range and noise tolerance.

\emph{Severity}: Medium. Cost and scalability are challenges, but innovative approaches such as Quantum Safe-as-a-Service (QaaS) models and hybrid networks are helping reduce deployment costs.

\emph{Timeline for Resolution}: Medium term (3–5 years). Market competition and advancements in off-the-shelf components are expected to make QKD increasingly affordable and scalable.

\item[\textbf{V.}] \textbf{Standardization and Interoperability}

The lack of standardized protocols and evaluation criteria poses a barrier to widespread adoption. However, organizations like ETSI, ISO, and ITU are actively developing global standards.

\emph{Severity}: Medium. Progress is steady, with global collaboration ensuring cross-vendor compatibility.

\emph{Timeline for Resolution}: Short to medium term (3–5 years). Certification frameworks are maturing rapidly and will soon establish clear interoperability guidelines.

\item[\textbf{VI.  }]   \hspace{0.05cm} \textbf{Integration with Classical Systems}

Integrating QKD with existing cryptographic frameworks and networks introduces complexity. However, hybrid systems combining QKD with classical encryption methods are showing promise.

\emph{Severity}: Low. Integration challenges are manageable with existing technology and ongoing developments in hybrid systems.

\emph{Timeline for Resolution}: Short term (2–3 years). Active development and testing are already underway.

\item[\textbf{VII.  }]  \hspace{0.15cm}  \textbf{Quantum Repeaters and Long-Distance Communication}

Transmission losses and the absence of practical quantum repeaters limit the achievable distance of QKD without trusted nodes. However, significant advancements in quantum memory and entanglement distribution are being made.

\emph{Severity}: Medium. While a challenge for global-scale QKD networks, near-term applications can rely on trusted nodes.

\emph{Timeline for Resolution}: Medium to long term (5–10 years). Progress in quantum repeaters and satellite-based QKD is accelerating, making this a solvable issue within the next decade.

\item[\textbf{VIII.  }]   \hspace{0.25cm} \textbf{Comparison with Post-Quantum Cryptography (PQC)}

While PQC provides an alternative to QKD for quantum-safe communication, it relies on computational assumptions. QKD offers the advantage of information-theoretic security based on physical principles.

\emph{Severity}: Low. PQC and QKD are complementary rather than competing technologies.

\emph{Timeline for Resolution}: Short term (1–2 years). Hybrid systems integrating QKD and PQC already provide practical and robust solutions.

\end{itemize}

Due to these trends, many countries and defense organizations prefer to monitor QKD's development and adopt it selectively or incrementally as the technology matures and its cost-effectiveness improves.

\subsection{Vision and Future Outlook}

Current QKD developments reflects the dual narrative of QKD’s transformative potential and the challenges limiting its scalability. Technological advancements in integrated photonics, cost-effective avalanche photodiodes, and continuous-variable QKD systems are driving down costs, making scalable implementations more feasible \cite{nphotonics}. Distance limitations are being addressed through satellite-based QKD, exemplified by China’s \textit{Micius} satellite enabling intercontinental secure communication, and the development of quantum repeaters to extend transmission ranges further \cite{satelliteqkd, repeaters}. Integration with classical cryptographic systems combines QKD’s information-theoretic security with traditional authentication and session management, ensuring compatibility with existing infrastructures. Government agencies, such as the UK’s NCSC and the USA’s NSA, emphasize the need for cost-benefit analyses given QKD’s high costs and scalability constraints, advocating for a cautious approach. In parallel, initiatives like NIST’s PQC standardization focus on scalable cryptographic alternatives. Despite this, private sectors, including finance, telecommunications, and technology, increasingly recognize QKD’s value. For instance, JPMorgan Chase has secured financial networks using QKD, BT has deployed quantum-secure industrial networks, and Toshiba has implemented QKD for healthcare data protection. These examples underscore QKD’s growing adoption in mission-critical applications as a complementary strategy to PQC.

\section{Recommendation}

Advancing QKD as a foundational technology for quantum-secure communication requires addressing challenges in implementation security, scalability, and interoperability. Governments and research institutions should establish national and regional QKD testbeds to integrate advanced protocols with existing systems, supporting real-world testing and standardization. Research into quantum repeaters and satellite-based QKD should be prioritized to overcome distance limitations, with international collaborations accelerating progress. Public-private partnerships can drive cost reduction, making QKD accessible for broader applications in enterprise and government use. Workforce development through education and training programs is essential to build expertise in quantum technologies, while active participation in global standardization efforts, such as those by ETSI and ISO, will ensure interoperability and facilitate widespread adoption. Together, these initiatives will position QKD as a critical solution to current and future cybersecurity challenges.
\section{Conclusion}

Quantum Key Distribution (QKD) is poised to become a cornerstone technology for securing communications in the quantum age, offering unparalleled information-theoretic security. Despite challenges in scalability, cost, and integration with classical systems, global investments underscore its strategic importance. Through focused research, innovation, and collaboration across industry and academia, countries and organizations worldwide can drive the adoption of QKD. By addressing key challenges and fostering international cooperation, QKD can become an integral component of the future cybersecurity landscape, ensuring robust protection against emerging quantum threats.

\section*{Biography}

{\bf Dr. Sebastian Kish} is a Research Scientist at CSIRO’s Data61, leading projects on Quantum Key Distribution since 2023. He specializes in developing and implementing QKD protocols for a quantum-encrypted network in Sydney. He holds a BSc (Hons) and PhD in Physics from The University of Queensland, with postdoctoral experience at UNSW and ANU in quantum communication and information. Sebastian has an extensive background in quantum physics and has published in top quantum journals.

\noindent
{\bf Dr. Josef Pieprzyk} (IACR Fellow, 2021) is a Professor at Institute of Computer Science, Polish Academy of Sciences and a Senior Principal Research Scientist at CSIRO’s Data61, focusing on cryptology and information security. His research includes cryptographic algorithms, secure protocols, and cybercrime prevention. He has authored 5 books, edited 10 conference proceedings, and published over 250 papers. He serves on editorial boards for journals like the \textit{International Journal of Information Security}.

\noindent
{\bf Dr. Seyit Camtepe} is a Principal Research Scientist at CSIRO’s Data61, leading the Autonomous and Software Security team. He focuses on innovative solutions to cybersecurity challenges, including Android malware and encryption. Seyit holds a PhD from Rensselaer Polytechnic Institute and has worked as a senior researcher at TU-Berlin and as an ECARD lecturer at QUT, Australia.

\bibliographystyle{unsrt}
\bibliography{myChapter_ref}


\end{document}